\documentclass{article}
\usepackage{emulateapj}
\usepackage{apjfonts}

\begin{document}

% Keep track of fig and table 
% numbering here because floats are not allowed in preprint format.
\newcommand{\figschem}{1}
\newcommand{\figlb}{2}

\newcommand{\tabphot}{1}

\submitted{To appear in The Astrophysical Journal Letters}

\title{Far-UV Emission from Elliptical Galaxies at $z=0.55^{1}$}

% USE FULL NAME 

\author{Thomas M. Brown$^2$, Charles W. Bowers, Randy A. Kimble}

\affil{Laboratory for Astronomy \& Solar Physics, Code 681, NASA/GSFC,
Greenbelt, MD 20771.\\   tbrown@pulsar.gsfc.nasa.gov,
bowers@band2.gsfc.nasa.gov, kimble@ccd.gsfc.nasa.gov.
}

\medskip

\author{Henry~C.~Ferguson}
\affil{Space Telescope Science Institute, 3700 San Martin Drive,
Baltimore, MD 21218.  ferguson@stsci.edu.}

\begin{abstract}

The restframe UV-to-optical flux ratio, characterizing the ``UV
upturn'' phenomenon, is potentially the most sensitive tracer of age
in elliptical galaxies; models predict that it may change by orders of
magnitude over the course of a few Gyr.  In order to trace the
evolution of the UV upturn as a function of redshift, we have used the
far-UV camera on the Space Telescope Imaging Spectrograph to image the
galaxy cluster CL0016+16 at $z=0.55$.  Our $25 \arcsec \times 25
\arcsec$ field includes four bright elliptical galaxies,
spectroscopically confirmed to be passively evolving cluster members.
The weak UV emission from the galaxies in our image demonstrates that
the UV upturn is weaker at a lookback time $\sim$5.6~Gyr earlier than our
own, as compared to measurements of the UV upturn in cluster E and S0
galaxies at $z=0$ and $z=0.375$.  These images are the first with
sufficient depth to demonstrate the fading of the UV upturn expected
at moderate redshifts.  We discuss these observations and the
implications for the formation history of galaxies.

\end{abstract}

\keywords{galaxies: evolution --- galaxies: stellar
content --- ultraviolet: galaxies}

\section{INTRODUCTION} \label{secintro}

Because they are composed of old, passively evolving populations,
elliptical galaxies offer great promise for tracing the evolution of
the Universe.  Whether elliptical galaxies formed through hierarchical
merging or monolithic collapse, one of the major goals in
extragalactic studies is the determination of the ``redshift of
formation,'' $z_F$, that marks the age where most of the stars in
early-type galaxies formed.  Recent studies of galaxy clusters out to
$z \sim 1$ indicate that most of the star formation had to be
completed at high redshift ($z \gtrsim 3$), followed by quiescent
evolution thereafter (Stanford, Eisenhardt, \& Dickinson
1998\markcite{SED98}; Kodama et al.\ 1998\markcite{KAB98}).

The UV upturn is a sharp rise in the spectra of E and S0 galaxies
shortward of restframe 2500~\AA.  It provides a sensitive tracer
of age for the oldest stars in these galaxies, and can potentially
constrain $z_F$.  Traditionally characterized by the $1550-V$ color,
the UV upturn in local galaxies originates in a population of hot
horizontal branch (HB) stars and their UV-bright progeny (see Brown et
al.\ 2000\markcite{B00} and references therein).  As first noted by
Greggio \& Renzini (1990\markcite{GR90}), the UV upturn should evolve
rapidly with age, through the evolution of HB morphology and the main
sequence turnoff mass.  Although all models of elliptical galaxy
evolution predict a rapid evolution in the UV upturn (e.g., Tantalo et
al.\ 1996\markcite{T96}), the timing for the UV upturn onset
depends strongly upon model parameters.

We have been undertaking a series of observations to trace the
evolution of the UV upturn as a function of redshift.  Faint Object
Camera (FOC) observations of Abell~370 ($z=0.375$) provided the first
detection of far-UV emission from quiescent elliptical galaxies above
$z = 0.05$ (Brown et al.\ 1998\markcite{BFDJ98}).  The very strong UV
emission found at $z=0.375$ suggests no evolution\\ 

{\small $^1$Based on observations with the NASA/ESA Hubble Space
Telescope obtained at the Space Telescope Science Institute, which is
operated by AURA, Inc., under NASA contract NAS~5-26555.

$^2$ NOAO Research Associate.}\\

\noindent 
in the UV upturn between our own epoch and one 4~Gyr earlier, a
finding inconsistent with some models of galaxy evolution, and
apparently consistent only for high values of $z_F$.  Here, we
describe observations that trace the UV upturn to higher redshift.  We
used the far-UV camera on the Space Telescope Imaging Spectrograph
(STIS) to measure the UV emission from four giant elliptical galaxies
in the cluster CL0016+16 at $z = 0.55$. Ground-based spectroscopy
confirms their passive evolution and cluster membership (Dressler \&
Gunn 1992\markcite{DG92}); morphological classification comes from
Wide Field Planetary Camera 2 (WFPC2) imaging in the F555W and F814W
bands (Smail et al.\ 1997\markcite{97}).  We assume the
currently popular cosmology of $\Omega_M = 0.3$, $\Omega_\Lambda =
0.7$, and $H_o = 67$ km s$^{-1}$ Mpc$^{-1}$, although our results are
more sensitive to $z_F$ than to the assumed cosmology.

\section{OBSERVATIONS} \label{secobs}

Using the STIS far-UV camera on 19 Aug and 21 Aug 1999, we observed a
$25\arcsec \times 25\arcsec$ field centered at
RA(J2000)=$0^h18^m33.2^s$ and Dec(J2000)=$\rm 16^o 26^\prime
9.7^{\prime\prime}$ in the cluster CL0016+16.  The total exposure was
27892 sec; 10 frames were taken in two visits, with dithering by 6
pixels to smooth out small-scale detector variations.  We used the
crystal quartz filter (F25QTZ) for a bandpass that spans
1450--2000~\AA, thus reducing the sky background from geocoronal
\ion{O}{1} and Lyman-$\alpha$ to negligible levels at little cost in
galaxy light, as the redshift ($z=0.55$) puts the Lyman limit at the
short wavelength cutoff.  Because the long wavelength cutoff of this
bandpass is due to detector sensitivity, red leak is also negligible;
we avoid the problematic red leak and red grating scatter that
hampered earlier attempts to measure the UV upturn at moderate
redshift (e.g., Windhorst et al.\ 1994\markcite{W94}).  The
photometric calibration is reliable at the 0.15 mag level (Baum et
al.\ 1998\markcite{B98}).  Recent efforts have revised the calibration
slightly; we assume the latest revision, with an accuracy that should
be better than 0.15 mag.  For reference, we assume that a flat
spectrum of $1.036\times 10^{-16}$ erg

\hskip -0.2in
\parbox{3.0in}{\epsfxsize=3.5in \epsfbox{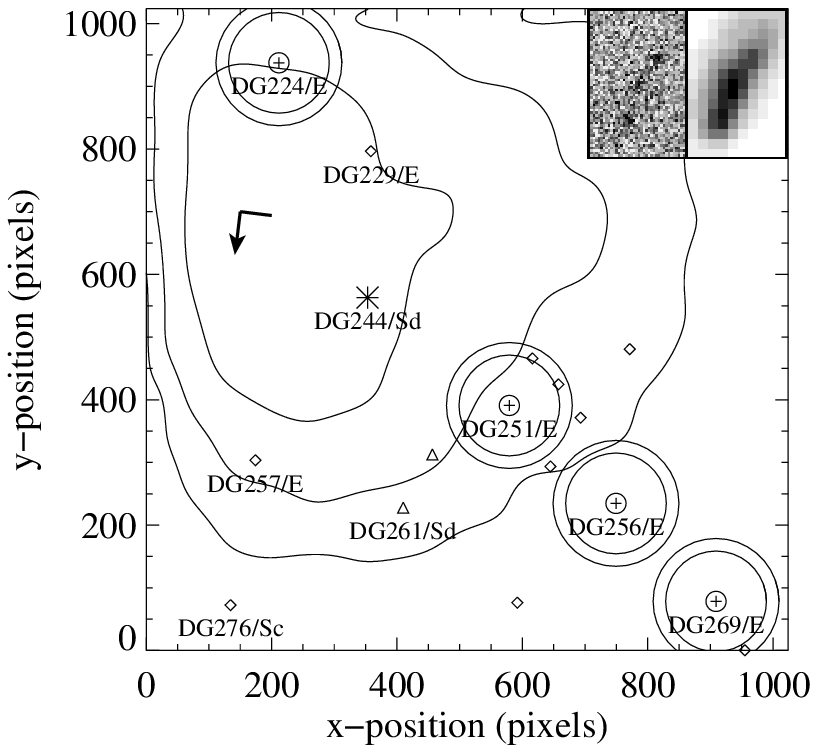}} \vskip 0.1in

%\begin{figure} \label{figschem}
\centerline{\parbox{3.5in}{\small {\sc Fig.~\figschem--} A schematic
of the STIS image.  Each bright cluster E galaxy is denoted by a
cross, photometric aperture, and sky annulus.  When there is a strong
dark background, its shape follows the contours shown (contour levels
are at $2\times$, $5\times$, and $10\times$ the lowest dark rate).  A
non-cluster spiral (asterisk) was used for astrometry and
registration; it is shown in the inset at $4\times$
magnification, as it appears in the STIS (left) and WFPC2 (right)
images.  Triangles denote $\sim 8 \sigma$ detections, and diamonds
mark other objects from the Smail et al.\ (1997) catalog at
$I_{814} < 24$ mag. Galaxies appearing in the Dressler \& Gunn
(1992\markcite{DG92}) catalog are labeled, along with their morphology
(Smail et al.\ 1997\markcite{S97}). }} \addtocounter{figure}{1}
%\end{figure} 

\medskip

\noindent
sec$^{-1}$ cm$^{-2}$ \AA$^{-1}$
produces one count per sec in the STIS bandpass.  A full description
of the instrument can be found in Woodgate et al.\
(1998\markcite{W98}) and Kimble et al.\ (1998\markcite{K98}).

Unfortunately, the mean temperature of the far-UV camera has increased
since STIS commissioning, and the dark background increases as a
function of temperature.  This increase appears as a ``glow'' that is
strongest in the upper left-hand quadrant of the detector
(Figure~\figschem), where the dark rate can be 20 times higher than
the nominal 6$\times 10^{-6}$ cts sec$^{-1}$ pix$^{-1}$.  The glow
varies from weak to strong in our 10 frames.

The Galactic foreground extinction toward our field was thought to be
$E(B-V) = 0.022$ mag at the time these observations were planned,
based upon \ion{H}{1} maps (Burstein \& Heiles 1984\markcite{BH84}).
The subsequent release of a new extinction map (Schlegel, Finkbeiner, \& Davis
1998\markcite{SFD98}), based upon an IRAS map of the Galactic dust,
revises the extinction estimate to 0.057 mag.  The increased dark
background and higher extinction both work to decrease the sensitivity
of our observations to far-UV emission, but do not preclude
constraints on the UV upturn.  When accounting for this extinction, we
assume the empirical model of Cardelli, Clayton, \& Mathis
(1989\markcite{CCM89}), which has been parameterized and tested from
the UV to IR.

Our STIS field includes four giant E galaxies (see Figure~\figschem);
three are detected at $>$2.5$\sigma$ significance.  We detect
three other objects from the Smail et al.\ (1997\markcite{S97}) catalog
at a significance of $\sim 8 \sigma$.  We use the bright background spiral,
DG244 ($z=0.6577$; Dressler \& Gunn 1992\markcite{DG92}), to confirm
the registration of our frames, and to determine the positions of the
fainter galaxies.  The accuracy of relative astrometry within WFPC2 or
STIS images is better than 0.1$\arcsec$, but the accuracy of absolute
astrometry can be worse than 1$\arcsec$, so the positions of objects
in the STIS field are determined relative to this bright spiral.  Most
of the objects in our chosen field are too red and faint to be
detected by STIS; their non-detection does not suggest an unexpected
lack of sensitivity, and all detected and non-detected objects are
consistent with the expected range of spectral energy distributions
(SEDs).  STIS repeatedly images UV-bright stars (e.g., globular
cluster NGC6681), and no drop in sensitivity is apparent in 
observations before and after our observation of the cluster
CL0016+16.

\section{DATA REDUCTION} \label{secred}

The STIS data was reduced via methods nearly identical to
those used for the Hubble Deep Field South, to which we refer
the interested reader for more details (Gardner et al.\
1999\markcite{G99}).  In brief, the images were processed via the
standard pipeline, excluding the dark subtraction, low-frequency flat
field correction, and geometric correction.  Dark frames from
July through August of 1999 were processed in the same manner, and
those with a strong glow ($> 2\times 10^{-5}$ cts sec$^{-1}$
pix$^{-1}$) were summed and fit with a cubic spline to produce a 
profile appropriate for our CL0016+16 observations (the shape of the
glow changes slowly with time, and so recent darks are required for
this fit).  A flat component and glow component to the dark background
were then subtracted from each CL0016+16 frame, and then the flat
field correction was applied.  The frames were registered by integer
shifts and summed via the DRIZZLE package.  The pixels in each frame
were weighted by the ratio of the exposure time squared to the dark
count variance, including a hot pixel mask.  The algorithm weights the
exposures by the square of the signal-to-noise (S/N) ratio for sources
that are fainter than the background, thus optimizing the summation to
account for the temporal and spatial variations in the dark
background.  The statistical errors in the final drizzled image (cts
pix$^{-1}$), for objects below the background, are given by the square
root of the final drizzled weights map scaled by the exposure time.
Geometric correction, which tends to smear out the pixels through
non-integer shifts, was not applied.  For small dithers and large
extraction apertures, it is not needed for object registration.

WFPC2 F814W data for the same field were obtained from the STScI
archive and reduced via standard techniques, including cosmic-ray
rejection and masking of problematic pixels.  The total exposure in
the F814W image is 16800 sec.  The summed image was used to determine
the restframe optical-band flux for each elliptical galaxy, and to
determine the relative positions of objects in the STIS field.

\section{PHOTOMETRY}
\label{secphot}

We performed aperture photometry on the far-UV frames using IDL,
taking a weighted average of the flux within a 16-pixel (0.4$\arcsec$)
radius, and a weighted average of the flux within a sky annulus of
radii 80 and 100 pixels.  The aperture includes the bright core of
each galaxy, as they appear in the WFPC2 data.  The aperture is small
enough to minimize the background in the far-UV measurement, but large
enough to avoid significant errors in the expected encircled energy
(due to the uncertainty in the position of the galaxies); it is also
relevant to measurements made of local galaxies, which use a nuclear
aperture (see \S \ref{secinterp}).  The weighting used the map of
statistical errors (see \S \ref{secred}), in the sense that pixels
were weighted less if they had less exposure (due to masked pixels) or
higher dark count rates.  This weighting did {\it not} weigh by counts
in the {\it data} frame, which would obviously bias the photometry
toward pixels with more source

\newpage

\parbox{5.5in}{
{\sc Table \tabphot:} Photometry

\begin{tabular}{|l|r|r|r|r|c|r|r|r|c|c|}
\tableline
&\multicolumn{2}{|c|}{position} & \multicolumn{2}{|c|}{catalog} &  & &\multicolumn{2}{|c|}{aperture phot.} & restframe & $1550-V$ \\  \cline{2-5} \cline{8-9}
&x   &     y & Smail$^a$ & DG$^b$ & Morph. &        & FUVQTZ       & F814W & $1550-V$ & 3$\sigma$ limit \\
&(pix)&(pix) & ID  & ID  & type$^a$ & z$^b$      & (cts)        & (DN) & (mag)  & (mag) \\
\tableline
           &212 &   938 & 650 & 224 & E/S0 & 0.5382 &  33$\pm$ 29 & 20577$\pm$94  & 3.7$_{-0.7}^{+2.3}$  &  2.3  \\   %2
Cluster    &579 &   391 & 724 & 251 & E    & 0.5433 &  61$\pm$ 22 & 34388$\pm$126 & 3.6$_{-0.3}^{+0.5}$  &  2.8  \\   %3
Elliptical &749 &   234 & 725 & 256 & E    & 0.5324 &  40$\pm$ 15 & 30450$\pm$111 & 4.0$_{-0.3}^{+0.5}$  &  3.1  \\   %4
Galaxies   &909 &    78 & 745 & 269 & E    & 0.5405 &  40$\pm$ 15 & 28964$\pm$79  & 3.9$_{-0.3}^{+0.5}$  &  3.1  \\   %5
\cline{2-11}
& \multicolumn{6}{|c|}{Sum of four cluster elliptical galaxies} & 174$\pm$43 & 114379$\pm 204$ & 3.8$_{-0.2}^{+0.3}$ & 3.2 \\
\tableline
\tableline
Other    &353 &   563 & 677 & 244 & Sd   & 0.6582 & 280$\pm$ 32 & 14693$\pm$67   &  N/A   &  N/A  \\   %1
Detected &410 &   228 & 705 & 261 & Sd   & ...    & 177$\pm$ 22  & 3831$\pm$69   &  N/A   &  N/A  \\   %14
Galaxies &456 &   313 & 696 & ... & ?    & ...    & 185$\pm$ 24  & 858$\pm$83    &  N/A   &  N/A  \\   %18
\tableline
\end{tabular}
$^a$ Smail et al.\ (1997\markcite{S97}).\\
$^b$ Dressler \& Gunn (1992\markcite{DG92}).\\
}

\noindent
counts.  We determined the local sky value from the mean instead of
the median in the sky annulus, because most pixels are ones and zeros.
Note that alternate methods for determining the sky background (e.g.,
fitting a surface to the local residual background) yield results well
within the 1$\sigma$ photometric errors.  The positions of objects in
the STIS image were determined from the geometrically-corrected WFPC2
F814W frame, relative to the spiral galaxy DG244 (see
Figure~\figschem).  Calculation of position in the STIS frame involves
a rescaling, rotation, translation, and geometric distortion, using
the geometric distortion coefficients of Malumuth
(1997\markcite{M97}).  We tested our positioning algorithm using WFPC2
and STIS images of the globular cluster NGC6681, which was observed
with STIS numerous times before and after our CL0016+16 observations
at various roll angles and positions.  Absolute positional accuracy
for objects in the STIS frame is 1--2 STIS pixels, well within the
16-pixel radius used for aperture photometry.

For the WFPC2 F814W frames, we performed aperture photometry with the
IRAF package PHOT, using a 4-pixel radius (0.4$\arcsec$), and
a sky annulus of radii 20 and 25 pixels, thus matching the photometry
done in the STIS image.  Note that this aperture size would produce
encircled energy agreement at the 5\% level for point sources, and
better agreement for extended sources (Robinson 1997\markcite{R97};
Holtzman et al.\ 1995\markcite{H95}), so the uncertainty in encircled
energy contributes less than 0.1~mag to $1550-V$.  We do not perform
an aperture correction because we are only interested in colors, not
the absolute fluxes.

Photometry was performed on all objects with $I_{814} < 24$ mag in the
Smail et al.\ (1997\markcite{S97}) catalog, plus a faint object
(\#696) that is obvious in the STIS image but faint in the WFPC2
frames.  Table~\tabphot\ gives the photometry for the four giant
elliptical galaxies, plus three objects detected at $\sim 8 \sigma$
in the STIS image.

\section{INTERPRETATION}
\label{secinterp}

The UV upturn is traditionally characterized by the restframe $1550-V$
color (see Burstein et al.\ 1988\markcite{B88}).  Conversion to
restframe $1550-V$ from our observed bandpasses depends upon the
assumption of an SED.  Brown et al.\ (1998\markcite{BFDJ98}) used the
spectra of three local elliptical galaxies (NGC1399, M60, and M49) to
convert their observed FOC bandpasses to restframe $1550-V$, and we
used the same spectra here.  We redshifted the spectra of NGC1399,
M60, and M49 to $z=0.55$, and then applied a foreground reddening of
$E(B-V)=0.057$ mag (Schlegel et al.\ 1998\markcite{SFD98}).  We then
used the IRAF CALCPHOT routine to calculate the STIS and WFPC2
countrates, giving: \vskip 0.05in

$1550-V = {\rm -2.5 \times log_{10}} (cps_{FUV}/dps_{F814W})-3.8$
mag, \\ \\ \vskip 2.25in

\noindent
where $cps_{FUV}$ is the STIS countrate (cts sec$^{-1}$), and
$dps_{F814W}$ is the WFPC2 countrate (DN sec$^{-1}$).  This conversion
gives the restframe $1550-V$ values shown in Table~\tabphot.  We also
calculate the $1550-V$ assuming the $3\sigma$ upper limit to the
far-UV flux.  Note that the $1550-V$ values in Table~\tabphot\ would
be redder if the foreground extinction is closer to $E(B-V)=0.022$ mag
(see \S \ref{secobs}), or if the far-UV SEDs are dominated by post-AGB
stars (see below).  The $1550-V$ color for the sum of the giant
elliptical galaxy photometry gives an average measurement of the UV
upturn in our sample.

Although we are looking to a significantly earlier epoch, the
assumption of a model SED over an empirical SED makes little
difference for the restframe $V$, because the observed F814W band
overlaps with restframe $V$.  For example, assuming that E galaxies
are $\sim$5~Gyr old at $z=0.55$, we calculated the
F814W$_{observed}-V_{restframe}$ color using the Bruzual \& Charlot
(1993\markcite{B93}) instantaneous burst SED of age 5~Gyr, and the
NGC1399 SED, assuming $z=0.55$ with a foreground reddening of
$E(B-V)=0.057$.  The difference in F814W$_{observed}-V_{restframe}$
for the two SEDs was only 0.02~mag.  The assumed SED has somewhat more
effect on the far-UV.  Our STIS bandpass spans restframe
935--1290~\AA, and the $1550-V$ color is based on the average flux
from 1250--1850~\AA.  However, we note that no model SED has been
tested in the far-UV for passively-evolving galaxies at moderate
redshift, and thus we prefer empirical SEDs.  Younger galaxies
should have a larger contribution from relatively hot post-AGB stars,
instead of HB stars (see Brown et al.\ 1997\markcite{BFDD97}), so our
conversion might systematically produce a bluer $1550-V$ than
the true restframe $1550-V$.  Alternatively, if younger galaxies
are dominated by even cooler HB stars than those in local galaxies,
we are calculating a redder $1550-V$ than the true value.

Another source of systematic error is the aperture size.  Local E
galaxies have been measured through 14$\arcsec$ diameter aperture (1.5
kpc at the distance of Virgo), sampling only the nuclear flux
(Burstein et al.\ 1988).  The $z=0.375$ measurements were made through
a 1.82$\arcsec$ diameter aperture (9.8 kpc at the distance of
Abell~370), including all of the galactic light detected by the FOC
(Brown et al.\ 1998).  We have measured the flux from our $z=0.55$
galaxies through a $0.8\arcsec$ diameter aperture (5.4 kpc at the
distance of CL0016+16), enclosing the light from the galaxy cores.
Our aperture is 3.6 times larger than that used for local galaxies,
but 1.8 times smaller than that used for the Abell 370 galaxies.  Ohl
et al.\ (1998) found that local E galaxies usually (but not always)
become redder in $1550-V$ at increasing radius; the color gradient is
not large (0.1--0.5 mag) over radii 7--25$\arcsec$, and the color of
the integrated light in an increasing aperture will change even less,
because most of the light comes from the central $\sim 2$ kpc even in
our larger apertures.  Thus, the aperture effects should be small, but
perhaps non-negligible ($\sim 0.1$ mag).

\section{DISCUSSION}

The average restframe $1550-V$ color for the four giant E
galaxies at $z=0.55$ is much redder than that observed
in clusters at $z=0$ and $z=0.375$.  At the $3\sigma$ upper limit to
the far-UV flux, none of the $z=0.55$ galaxies are as blue as the
strong UV-upturn galaxies observed locally (e.g., M60 or NGC1399).
Ideally, one would want to compare galaxies at similar velocity
dispersion ($\sigma_v$) in each epoch, because $\sigma_v$ can be
measured in a model-independent manner, and because the UV upturn
strongly correlates with $\sigma_v$ locally (Burstein et
al.\ 1988). Measurements of $\sigma_v$ are unavailable for some of the
$z=0.375$ galaxies and all of the $z=0.55$ galaxies, but these
galaxies were selected from the brightest and largest in each cluster,
and would likely show strong UV upturns if observed locally.  The
faint far-UV emission from the $z=0.55$ galaxies thus demonstrates that the
UV upturn is sensitive to age.  We plot these UV upturn measurements
in Figure~\figlb, along with the models of Tantalo et al.\
(1996\markcite{T96}), which trace the chemical evolution of elliptical
galaxies under the assumption of gas infall.  Note that the scatter in
local UV upturn measurements is affected by the wide range of sizes
and luminosities in the sample; the moderate-redshift samples are more
homogeneous, but the measurements themselves are less statistically
significant.

In the Tantalo et al.\ (1996\markcite{T96}) models, the UV upturn
appears at an age of $\sim 6.5$ Gyr, becomes quite blue by 9~Gyr, and
is flat thereafter.  Assuming the models are correct and that the
oldest stars in cluster E galaxies all formed at a common
redshift, our measurements imply $z_F \sim 4$.  If elliptical galaxies
form through hierarchical merging, the age of the oldest stars may
predate such merging; alternatively, the age of the oldest stars may
reflect the age of the galaxies themselves, if they were assembled
through monolithic collapse.  However, the Tantalo et al.\
(1996\markcite{T96}) models rely on several parameters (e.g., time of
onset for galactic winds, efficiency of the star formation rate,
accretion timescale, etc.) that can be tuned to delay or accelerate
the onset of the UV upturn.  In this sense, our results say more about
the evolution of HB morphology than $z_F$; if galaxies form their
oldest stars at $z > 4$, our data imply that a blue HB population
cannot arise until ages greater than 7 Gyr.

Although the elliptical galaxies in our program are chosen through
ground-based spectroscopy that confirms their passive evolution, the
strong UV upturn found at $z=0.375$ could theoretically be due to
residual star formation instead of evolved populations.  This star
formation would have to cease by the present epoch, because the UV
emission from local cluster elliptical galaxies is clearly due to HB
stars instead of star formation.  If all clusters are alike, our new
$z=0.55$ observations reinforce the case against star formation at
$z=0.375$, because such star formation would likely be even stronger
at a larger lookback time.  Measurements at $z=$ 0, 0.375, and 0.55
demonstrate that the UV upturn rises sharply over a few Gyr, and then
levels off.  This line of inquiry into the evolution of galaxies is
still in the early stages, and further observations are needed over a
range of redshifts.  Such measurements should become much more
feasible in HST Cycle 10, once a cooler is connected to STIS (reducing
the dark background to minimal levels).  

\smallskip

\hskip -0.2in
\parbox{3.5in}{\epsfxsize=3.5in \epsfbox{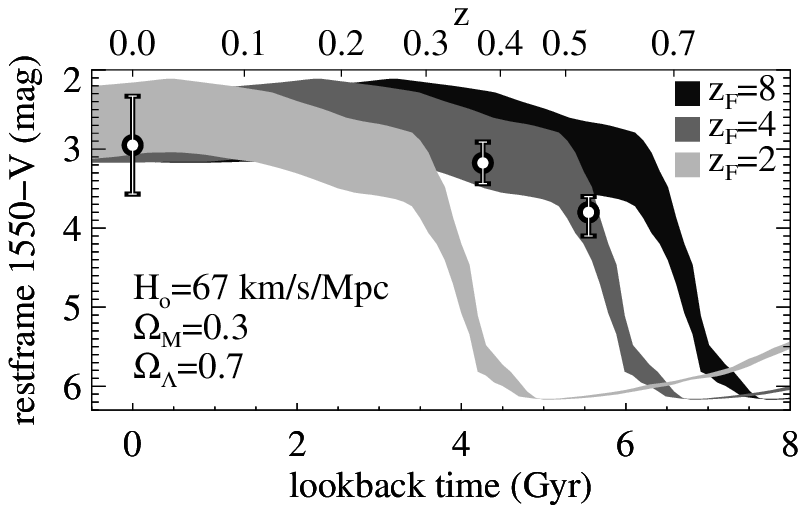}} \\

%\begin{figure} \label{figschem}
\centerline{\parbox{3.5in}{\small {\sc Fig.~\figlb--} Evolution of the
restframe $1550-V$ color according to the models of Tantalo et al.\
(1996\markcite{T96}) (shaded), viewed as a function of lookback time,
assuming a reasonable set of cosmological parameters and three
different epochs of galaxy formation (labeled).  The spread in
$1550-V$ is bounded by models at $M=3\times 10^{12}$ and $M=10^{12}
M_{\odot}$.  Note the sudden onset of UV emission caused by the
appearance of metal-rich hot HB stars.  The points represent the mean
$1550-V$ color measured for quiescent E and S0 galaxies in clusters at
$z=0$ (Virgo, Fornax, and Coma; Burstein et al.\ 1988), $z=0.375$
(Abell 370; Brown et al.\ 1998), and $z=0.55$ (CL0016+16). The error
bars at $z=0$ and $z=0.375$ give the rms in galaxy colors
(photometric errors are negligible); the error bar at
$z=0.55$ gives the photometric error for the summed galaxies 
(rms in the galaxy colors is smaller).}}  \addtocounter{figure}{1}
%\end{figure} 

\acknowledgments Support for this work was provided by NASA through
the STIS GTO team funding.  TMB acknowledges support at GSFC by
NAS~5-6499D.  We wish to thank the MORPHS project for making their
cluster data and catalogs publicly available.

\end{document}